\documentclass[12pt, epsfig]{article}
\usepackage{graphicx}
  \usepackage{amsthm,amsfonts}
  \usepackage{amsmath}
\newcommand{\bea}   {\begin{eqnarray}}
\newcommand{\eea}   {\end{eqnarray}}

\begin{document}
\renewcommand{\thefootnote}{\fnsymbol{footnote}}

\thispagestyle{empty}

\title{The quasi-nonassociative exceptional $F(4)$ \\deformed quantum oscillator}

\author{N. Aizawa\thanks{{E-mail: {\em aizawa@p.s.osakafu-u.ac.jp}}},\quad 
Z. Kuznetsova\thanks{{E-mail: {\em zhanna.kuznetsova@ufabc.edu.br}}}
\quad and\quad F.
Toppan\thanks{{E-mail: {\em toppan@cbpf.br}}}
\\
\\
}
\maketitle

\centerline{$^{\ast}$
{\it Department of Physical Science, Graduate School of Science,}}

{\centerline{\it\quad Osaka Prefecture University, Nakamozu Campus, Sakai, Osaka 599-8531 Japan.}}
\centerline{$^{\dag}$
{\it UFABC, Av. dos Estados 5001, Bangu,}}{\centerline {\it\quad
cep 09210-580, Santo Andr\'e (SP), Brazil.}
\centerline{$^{\ddag}$
{\it CBPF, Rua Dr. Xavier Sigaud 150, Urca,}}{\centerline {\it\quad
cep 22290-180, Rio de Janeiro (RJ), Brazil.}
~\\
\maketitle
\begin{abstract}
We present the deformed (for the presence of Calogero potential terms) one-dimensional quantum oscillator with the exceptional Lie superalgebra $F(4)$ as \\spectrum-generating superconformal algebra. The Hilbert space is given by a $16$-ple of square-integrable functions. The energy levels are $\frac{2}{3}+n$, with $n=0,1,2,\ldots$.  The ground state is $7$ times degenerate. The excited states are $8$ times degenerate.  The $(7,8,8,8,\ldots )$
semi-infinite tower of states is recovered from the $(7;8;1)$ supermultiplet of the ${\cal N}=8$ worldline supersymmetry. The model is unique, up to similarity transformations, and admits an octonionic-covariant formulation which manifests itself as ``quasi-nonassociativity". This means, in particular, that the Calogero coupling constants are expressed in terms of the octonionic structure constants.\par
The associated $F(4)$ superconformal quantum mechanics is also presented.

~\\\end{abstract}
\vfill

\rightline{CBPF-NF-002/17
}

\newpage
\section{Introduction}
We present the unique (up to similarity transformations) one-dimensional superconformal quantum mechanical system with
the $F(4)$ Lie superalgebra as dynamical symmetry; its associated (via the de Alfaro-Fubini-Furlan construction \cite{dff}) deformed (due to the presence of Calogero's \cite{cal} potential terms) quantum oscillator, is also derived. The Hilbert space of the latter model and its discrete, bounded from below, spectrum is obtained.\par
The division algebra of the octonions are at the core of several exceptional structures in mathematics. The $5$ exceptional Lie algebras are all related to the octonions. Indeed $g_2$ is the Lie algebra of the $G_2$ group of automorphisms of the octonions, while the remaining $4$ exceptional Lie algebras, $f_4$, $e_6$, $e_7$ and $e_8$,
are induced by the octonions via the Freudenthal-Tits magic square construction. About simple Lie superalgebras, the exceptional Lie superalgebras $G(3)$ and $F(4)$ entering the Kac's classification \cite{kac} admit an octonionic realization \cite{sud}.\par
We demanded octonionic covarariance as a key ingredient to derive the $F(4)$ models. This means that the differential operators of the spectrum-generating superalgebra are expressed in terms of the octonionic structure constants.\par
We obtained at first the most general, octonionic covariant and scale invariant, ${\cal N}=8$ supersymmetric quantum mechanics defined on $8$ bosonic and $8$ fermionic fields. We further implemented the superconformal constraint which selects a critical theory with enhanced symmetry. The existence, at the critical value, of $7$ linear constraints for the $28$ $R$-symmetry generators, unambiguously determines
$F(4)$ (whose $R$-symmetry is $so(7)$, with $21=28-7$ generators) as the dynamical symmetry superconformal algebra of the model.\par
The model is characterized by ``quasi-nonassociativity". The Calogero's coupling constants are expressed
in terms of the octonionic structure constants $C_{ijk}$ which encode the nonassociativity of the octonions.\par
Octonionic covariance was used in \cite{kuroto} to construct the most general classical ${\cal N}=8$ extended supersymmetric one-dimensional sigma-model in the Lagrangian framework based on the $(1;8;7)$ irreducible supermultiplet, see \cite{pato}, of the ${\cal N}=8$ worldline supersymmetry.
 It was proved in \cite{di} that an $F(4)$ superconformally invariant, one-dimensional sigma-model results from a suitable restriction of the parameters entering the ${\cal N}=8$, $(1;8;7)$ theory. In \cite{kuto} and 
\cite{khto} the systematic construction of one-dimensional superconformal algebras in terms of worldline supermultiplets ($D$-module representations) was obtained. We recall that one-dimensional superconformal algebras are simple Lie superalgebras satisfying a class of restrictions, resulting from their possible interpretation as dynamical symmetries of superconformal mechanics. Here it is sufficient to remind that the even sector should be decomposed into a direct sum $sl(2)\oplus R$, with $R$ known as the $R$-symmetry algebra, while the odd sector is decomposed into a set of ${\cal N}+{\cal N}$, dually related generators.\par
From \cite{{kuto},{khto}} follows, in particular, that octonionic-covariant $D$-module realizations are encountered at ${\cal N}=7$  for the exceptional $G(3)$ superalgebra and at ${\cal N}=8$ for the superalgebras $D(4,1)\approx osp(8|2)$ and $F(4)$. \par
It was noted in \cite{hoto} that ``trigonometric" $D$-module representations allow the construction of worldline sigma-models which correspond to a classical (and superconformal) version of  the de Alfaro-Fubini-Furlan prescription.\par
The quantization of the interacting trigonometric models was first obtained in \cite{cht}, by applying standard techniques to pass from the classical Lagrangian to the quantum Hamiltonian formulation. In particular
deformed quantum oscillators with $D(2,1;\alpha)$ as spectrum-generating superalgebras were derived.\par
The method used in this paper allows to directly construct a quantum mechanical system, bypassing the
scheme of deriving at first  a classical Lagrangian model  which is later quantized in the Hamiltonian framework. For this reason we did not need to introduce a classical Lagrangian trigonometric version (which has not been constructed, yet)  of the
classical $F(4)$ one-dimensional sigma-model introduced in  \cite{di}.\par

We postpone to the Conclusions a further discussion about the results of our model and about some issues of nonassociativity in physics.

\par
 
The scheme of the paper is as follows. In Section {\bf 2} we introduce the octonionic-covariant formulation. In Section {\bf 3} we derive the octonionic-covariant, scale invariant,  ${\cal N}=8$ supersymmetric quantum mechanics. In Section {\bf 4} we derive the differential realization of the $F(4)$ Lie superalgebra. The deformed $F(4)$ oscillator is introduced in Section {\bf 5}. In Section {\bf 6} the spectrum of the theory is obtained and the quasi-nonassociativity of the model is discussed. In the Conclusions we comment about our results and the issue of nonassociativity in physics.

\section{Octonions and the octonionic covariance}

The octonionic multiplication is encoded in the basic relations, for the seven imaginary octonions $e_i$
($i=1,2,\ldots,7$),
\bea\label{octonions}
e_i e_j &=& -\delta_{ij}+C_{ijk}e_k.
\eea
Here and in the following, unless otherwise specified, the sum over repeated indices is understood. $C_{ijk}$ is the totally
antisymmetric octonionic structure constant. Besides the rank $3$ $C_{ijk}$ tensor, two more totally antisymmetric
constant tensors (of rank $4$ and $7$) are compatible with the octonionic multiplication; they are given by
$C_{ijkl}$ and $\epsilon_{ijklmnp}$. We assume the  rank $3,4,7$ totally antisymmetric tensors to be normalized according to:
\bea
&C_{123}=C_{147}=C_{165}= C_{246}=C_{257}=C_{354}=C_{367}=1,&\nonumber\\
&C_{4567}=C_{2356}=C_{2437}= C_{1357}=C_{1346}=C_{1276}=C_{1245}=1,&\nonumber\\
&\epsilon_{1234567}=1.&
\eea
Due to a relation involving the three totally antisymmetric constant tensors,  only two of them are independent. The relation can be expressed as
\bea
6 C_{ijkl} &=& \epsilon_{ijklmnp}C_{mnp}.
\eea
The seven imaginary octonions can be conveniently arranged in the famous Fano's projective plane (see \cite{bae} for a review). The non-vanishing $C_{ijk}$'s correspond to three points belonging to one of its seven lines. The non-vanishing $C_{ijkl}$'s correspond to the four points which are complementary to each one of the seven lines.\par
The octonions induce a realization of the Clifford algebra $Cl(0,7)$ based on the following construction. Let $x=x_0+x_je_j$ be a real octonion, parametrized by eight real numbers $x_0, x_j\in {\mathbb R}$, so that ${\vec x}=(x_0,x_j)^T$ is an $8$-component real vector. The mapping
\bea\label{octmap}
x&\mapsto& e_ix = x'_i
\eea
can be expressed as a linear transformation
\bea
{\vec x'_i} &=&\gamma_i {\vec x}.
\eea
It is easily proved that the seven $8\times 8$ matrices $\gamma_i$ so induced satisfy the $Cl(0,7)$ Clifford algebra fundamental
relation
\bea\label{cl07}
\gamma_i\gamma_j+\gamma_j\gamma_i&=& -2\delta_{ij} {\mathbb I}_8,\quad\quad i,j=1,2,\ldots, 7
\eea
(here and in the following ${\mathbb I}_n$ denotes the $n\times n$ identity matrix).\par
Furthermore, their entries are expressed in terms of the octonionic structure constants $C_{ijk}$ according to
\bea\label{gmatrices}
({\gamma_i})_{LM} &=& \left(\begin{array}{c|c}0&\delta_{im}\\ \hline
-\delta_{il}&C_{ilm}\end{array}\right),
\eea 
where $L$, $M$ take values $L=0,l$ and $M=0,m$, with $l,m=1,2,\dots, 7$. The (\ref{gmatrices}) matrices are obtained, up to an overall sign,  from the (\ref{octmap}) map.\par
The double role played by $C_{ijk}$ should be duly noted. Entering (\ref{octonions}) it is responsible for the
non-associativity of the octonionic multiplication. We have, e.g.:
\bea
(e_1e_2)e_4= e_3e_4=-e_5 &\neq & e_1(e_2e_4)=e_1e_6=e_5.
\eea
On the other hand $C_{ijk}$ enters (\ref{cl07}) as well, providing a matrix realization for the associative $Cl(0,7)$
Clifford algebra. One can say that, in this matrix realization, $C_{ijk}$ encodes the remnant of the non-associativity
of the octonions.\par
The $Cl(0,7)$ Clifford algebra gives a basis for the $64$-dimensional vector space of $8\times8$ real matrices. Schematically, the elements of given rank $r=0,1,2,3$ (entering $\gamma^{(r)}$) are 
\bea\label{rank7}
\gamma^{(0)}& \equiv& {\mathbb I}_8,\nonumber\\
\gamma^{(1)}&\equiv& \gamma_i,\nonumber\\
\gamma^{(2)} &\equiv & \gamma_i\gamma_j \quad (i<j),\nonumber\\
\gamma^{(3)} &\equiv & \gamma_i\gamma_j\gamma_k\quad (i<j<k).
\eea
Due to Hodge duality, the product of $7-r$ different matrices $\gamma_i$ is equivalent to the product of matrices of rank $r$.
There is a total number of $\left(\begin{array}{c} 7\\r\end{array}\right)$ matrices of rank $r$, so that
\bea
\left(\begin{array}{c} 7\\0\end{array}\right)+\left(\begin{array}{c} 7\\1\end{array}\right)+\left(\begin{array}{c} 7\\2\end{array}\right)+\left(\begin{array}{c} 7\\3\end{array}\right)&=&1+7+21+35=64.
\eea
$\gamma^{(0)}$ and $\gamma^{(3)}$ provide the basis for the $36$ symmetric $8\times 8$ matrices, while 
$\gamma^{(1)}$ and $\gamma^{(2)}$ provide the basis for the  $28$ antisymmetric $8\times 8$ matrices.\par
The introduction of supersymmetry requires acting on a ${\mathbb Z}_2$-graded vector space of even and odd elements (identified with bosons and fermions), with block-antidiagonal supersymmetry operators. In the present case this can be obtained by doubling the size of the vector space by introducing the nine $16\times16$ matrix generators of the $Cl(9,0)$ Clifford algebra $\Gamma_A$ ($A=1,2,\ldots, 8, 9$) through the positions
\bea
&\Gamma_i = \left(\begin{array}{cc}0&\gamma_i\\ 
-\gamma_i&0\end{array}\right),\quad
\Gamma_8 = \left(\begin{array}{cc}0&{\mathbb I}_8\\ 
{\mathbb I}_8&0\end{array}\right),\quad
\Gamma_9 = \left(\begin{array}{cc}{\mathbb I}_8&0\\ 
0&-{\mathbb I}_8\end{array}\right).&
\eea
In the following different symbols are employed for different ranges of values: $i,j=1,2,\ldots, 7$, while $I,J=1,2,\ldots, 8$ and $A,B=1,2,\ldots 9$.\par
We have, by construction,
\bea
\Gamma_A\Gamma_B+\Gamma_B\Gamma_A&=& 2\delta_{AB} {\mathbb I}_{16}.
\eea
The block-diagonal matrix $\Gamma_9$ can now be identified with the Fermion Parity Operator of the Supersymmetric Quantum Mechanics. Its $\pm 1$ eigenvalues determine the $8$-dimensional bosonic ($+1$) and fermionic ($-1$) vector spaces.\par
The $256$-dimensional vector space of  $16\times 16$ real matrices can be expressed in terms of the $Cl(9,0)$ Clifford algebra matrix generators $\Gamma_A$'s. The  different $\left(\begin{array}{c} 9\\{\bf r}\end{array}\right)  
$ rank ${\bf r}$ tensors are compactly written as $\Gamma^{({\bf r})}\equiv \Gamma_{A_{1}}\ldots \Gamma_{A_{\bf r}}$ for $A_1<A_2<\ldots <A_{\bf r}$ (${\bf r}=0,1,2,3,4$). The analogue of formula (\ref{rank7}) now reads
\bea\label{rank9}
\left(\begin{array}{c} 9\\0\end{array}\right)+\left(\begin{array}{c} 9\\1\end{array}\right)+\left(\begin{array}{c} 9\\2\end{array}\right)+\left(\begin{array}{c} 9\\3\end{array}\right)+
\left(\begin{array}{c} 9\\4\end{array}\right)&=&1+9+36+84+126=256.
\eea
Due to the different status of $\Gamma_9$ (which is block-diagonal) with respect to the $8$ remaining $\Gamma_I$'s (which are block-antidiagonal), a more refined decomposition singles out $\Gamma_9$. A further refinement singles out $\Gamma_8$ (which is scalar with respect to the octonionic imaginary index) from the seven $\Gamma_i$'s which carry the vectorial index $i$ associated to the imaginary octonions. Taking into account this $7+1+1$ decomposition, we arrive at the following table. The second column denotes the block diagonal ($dg$) versus the block antidiagonal ($ad$) character of the matrices; the third column denotes their symmetry ($SYM$) versus antisymmetry ($AS$) property; the last column indicates their total number $N_b$. The symbol $\Gamma^{(r)}$ (for $r$ not in boldface font) denotes $\Gamma^{({r})}\equiv \Gamma_{i_{1}}\ldots \Gamma_{i_{r}}$ for $i_1<i_2<\ldots <i_{ r}$, with $i_r$ ranging from $1$ to $7$.
We have
\bea \label{table90}& \begin{array}{|c|c|c|c|}\hline
&dg/ad&SYM/AS&N_b\\\hline
{\mathbb I}&dg&SYM&1 \\ \hline
\Gamma^{(1)} &ad&SYM&7 \\
\Gamma_8 &ad&SYM&1\\
\Gamma_9 &dg&SYM&1\\ \hline
\Gamma^{(2)}&dg&AS&21\\
\Gamma^{(1)}\Gamma_8 &dg&AS&7\\
\Gamma^{(1)}\Gamma_9 &ad&AS&7\\
\Gamma_8\Gamma_9 &ad&AS&1\\ \hline
\Gamma^{(3)}&ad&AS&35\\
\Gamma^{(2)}\Gamma_8 &ad&AS&21\\
\Gamma^{(2)}\Gamma_9 &dg&AS&21\\
\Gamma^{(1)}\Gamma_8\Gamma_9 &dg&AS&7\\ \hline
\Gamma^{(4)}&dg&SYM&35\\
\Gamma^{(3)}\Gamma_8 &dg&SYM&35\\
\Gamma^{(3)}\Gamma_9 &ad&SYM&35\\
\Gamma^{(2)}\Gamma_8\Gamma_9 &ad&SYM&21\\ \hline
\end{array}&
\eea
\section{Scale-invariant ${\cal N}=8$ Supersymmetric Quantum Mechanics}

We are now in the position to introduce the ${\cal N}=8$ Supersymmetric Quantum Mechanics, defined by the
(anti)commutators
\bea\label{sqm}
\{ Q_I, Q_J\} &=& 2\delta_{IJ} H,\nonumber\\
\relax [H,Q_I]&=& 0.
\eea
The eight supersymmetry operators $Q_I$ are Hermitian and block-antidiagonal. $H$ is a Hamiltonian which can
be expressed as
\bea
H&=& -\frac{1}{2}\partial_x^2{\mathbb I} +V(x),
\eea
 where $V(x)$ is a block-diagonal, real, symmetric matrix potential.\par
Since we are interested in Superconformal Quantum Mechanics (and the oscillator models possessing the associated superalgebra as dynamical symmetry), we investigate at first the condition to obtain scale-invariant Supersymmetric
Quantum Mechanics. It follows, in particular, that the potential $V(x)$ should be expressed as
\bea
V(x) &=& \frac{1}{x^2} V,
\eea
where $V$ is a block-diagonal, constant, symmetric ($V^T=V$) matrix.\par
For the moment we are working with $16\times 16$ real matrices. It makes sense to express the differential 
part entering $Q_I$'s to be proportional to $\Gamma_I\Gamma_9\partial_x$.  This set of operators indeed satisfies the requirements of being Hermitian and block-antidiagonal. Concerning the potential terms, proportional to $\frac{1}{x}$, we assume the most general ones satisfying the criteria of being block-antidiagonal, symmetric and octonionic covariant.  Under these assumptions we propose the following Ansatz for $Q_8$ and $Q_i$'s:
\bea\label{qs}
Q_8 &=&\frac{1}{\sqrt 2}\left( \Gamma_8\Gamma_9\partial_x +\frac{1}{x}E_8\right),\quad E_8=a C_{ijk}\Gamma_i\Gamma_j\Gamma_k\Gamma_9+b\Gamma_8,\nonumber\\
Q_i &=& \frac{1}{\sqrt 2}\left(\Gamma_i\Gamma_9\partial_x+ \frac{1}{x}E_i\right), \quad E_i= c C_{ijk}\Gamma_j\Gamma_k\Gamma_8\Gamma_9 + dC_{ijkl} \Gamma_j\Gamma_k\Gamma_l\Gamma_9+ e\Gamma_i,
\eea
where the constants  $a,b,c,d,e$ are real coefficients to be determined by the closure of  (\ref{sqm}).\par
The requirement $\{Q_i,Q_j\} =0$ for $i\neq j$ is solved for either
\bea\label{first}
d=\frac{1}{3}c, &&e=-\frac{1}{2} +6c,
\eea
 or
\bea\label{second}
d=-\frac{1}{3}c, && e=\frac{1}{2} +6c.
\eea
Both restrictions produce a Hamiltonian $H$, invariant under ${\cal N}=7$ supersymmetries, possessing a diagonal
potential.  The constant matrix $V = diag(v_1,v_2,v_3,\ldots, v_{16})$ is given, in case (\ref{first}) by
\bea\label{Vcase1}
&v_1=\ldots=v_8=-\frac{1}{8}+32c^2, \quad v_9= \frac{3}{8}+8c+32c^2,\quad v_{10}=\ldots =v_{16}=\frac{3}{8}-8c+32c^2.\quad&
\eea
In case (\ref{second}) $V$ is given by
\bea\label{Vcase2}
&v_1=\frac{3}{8} -8c+32c^2,\quad v_2= \ldots=v_8= \frac{3}{8}+8c+32c^2, \quad v_9=\ldots =v_{16} = -\frac{1}{8}+32c^2.\quad&
\eea
A scale-invariant ${\cal N}=8$ Supersymmetric Quantum Mechanics is recovered by further setting,
in both cases, 
\bea\label{third}
a=-\frac{1}{3}c,&& \quad b=e.
\eea
The scale-invariant Hamiltonians $H$, with matrix potentials determined by either (\ref{Vcase1}) or (\ref{Vcase2}), are ${\cal N}=8$ supersymmetric. They depend on an arbitrary real coupling constant $c$. 

\section{The superconformal algebra realization}

The introduction of a superconformal algebra requires the presence of eight operators ${\widetilde Q}_I$, the superconformal partners of the supercharges $Q_I$. They have to be block-antidiagonal and, for dimensional reason,
proportional to $x$. The anticommutators $\{Q_I, {\widetilde Q}_J\}$ should produce the dilatation operator $D$
(from $I=J$) and the $R$-symmetry generators (from $I\neq J$). The dilatation operator $D$ should contain a term proportional to $x\partial_x{\mathbb I}$. This requirement rules out the most natural choice for ${\widetilde Q}_I$, consisting in setting
${\widetilde Q}_I\propto x\Gamma_I$. The anti-Hermitian choice ${\widetilde Q}_I\propto x\Gamma_I\Gamma_9$,
on the other hands, nicely works. In order to have Hermitian operators ${\widetilde Q}_I$, we therefore need to introduce the
imaginary unit $i$. It follows that, while scale-invariant ${\cal N}=8$ Supersymmetric Quantum Mechanics can be realized on $8$ real bosonic and $8$ real fermionic fields, in order to have ${\cal N}=8$ Superconformal Quantum Mechanics, we need to double (counting in real components) the number of fields, acting on $8$ complex bosonic
and $8$ complex fermionic fields.\par
Conveniently normalized, the operators ${\widetilde Q}_I$'s are chosen to be
\bea
{\widetilde Q}_I &=& \frac{i}{\sqrt 2} x\Gamma_I\Gamma_9.
\eea
They satisfy the anticommutators
\bea
\{{\widetilde Q}_I, {\widetilde Q}_J\} &=& 2\delta_{IJ} K,
\eea
where
\bea\label{K}
K&=& \frac{1}{2}x^2{\mathbb I}_{16}.
\eea
The next topic consists in investigating the anticommutators $\{Q_I, {\widetilde Q}_J\}$ and determinining under
which conditions (if any) an ${\cal N}=8$ superconformal algebra is recovered. \par
At any given $I$ we have 
\bea
\{Q_I, {\widetilde Q}_I\} &=& -i(x\partial_x+\frac{1}{2}){\mathbb I}_{16}:= D.
\eea
Therefore, without loss of generality, we can set
\bea\label{R}
\{Q_I, {\widetilde Q}_J\} &=& \delta_{IJ} D + R_{IJ},
\eea
where, for $I\neq J$,
\bea
R_{IJ} &=&\frac{i}{2}(-\Gamma_I\Gamma_J+\{E_I,\Gamma_J\Gamma_9\}).
\eea
We recall that the operators $E_I$ have been introduced in (\ref{qs}). Either the choice (\ref{first},\ref{third}) or  the choice (\ref{second},\ref{third}) of the real parameters have been  assumed.
For both choices, the $R_{IJ}$'s turn out to be  antisymmetric with respect to the $I\leftrightarrow J$ exchange:
\bea
R_{IJ}&=& - R_{JI}. 
\eea
In order to guarantee the closure, as Lie superalgebra, of the set of generators 
\\$H,D,K, Q_I, {\widetilde Q}_I, R_{IJ}$,
we need to check under which condition the $R_{IJ}$'s operators form a closed ($R$-symmetry) Lie algebra and the fermionic operators $Q_I$'s
(${\widetilde Q}_I$'s) belong to an $R$-symmetry representation. \par
These two requirements select specific values for $c$ entering (\ref{qs}).
With the choice (\ref{first}) for $d,e$, the value of $c$ is fixed to be $c=\frac{1}{12}$. Taking into account (\ref{third}),
the parameters $a,b,c,d,e$ entering (\ref{qs}) are thus determined to be
\bea\label{one}
&a=-d=-\frac{1}{36}, \quad c=\frac{1}{12},\quad b=e=0.&
\eea
With the choice (\ref{second}) for $d,e$, the selected value of $c$ is $c=-\frac{1}{12}$. The parameters 
$a,b,c,d,e$ entering (\ref{qs}) are given in this case by
\bea\label{two}
&a=d=\frac{1}{36}, \quad c=-\frac{1}{12}, \quad b=e=0.&
\eea
The two choices turn out to be equivalent. In the first case the potential is expressed, see (\ref{Vcase1}), by
\bea\label{vcase1}
&v_1=\ldots=v_8=\frac{7}{72}, \quad v_9= \frac{91}{72},\quad v_{10}=\ldots =v_{16}=-\frac{5}{72}.\quad&
\eea
In the second case, see (\ref{Vcase2}), the potential is expressed by 
\bea\label{vcase2}
&v_1=\frac{91}{72},\quad v_2= \ldots=v_8= -\frac{5}{72}, \quad v_9=\ldots =v_{16} = \frac{7}{72}.\quad&
\eea
If we express the first Hamiltonian $H_1$, in equal blocks decomposition, as
\bea
H_1 &=& \left(\begin{array}{cc} H_B&0\\0&H_F\end{array}\right),
\eea
the second Hamiltonian $H_2$ is obtained by exchanging bosons with fermions through a similarity transformation
induced by $\Gamma_8$:
\bea
H_2 &=& \left(\begin{array}{cc} H_F&0\\0&H_B\end{array}\right) = \Gamma_8 H_1 \Gamma_8.
\eea
Any operator $g_{2}$ obtained from the $c=-\frac{1}{12}$ choice is related to an operator $g_1$ obtained from the 
$c=\frac{1}{12}$ choice via the similarity transformation
\bea
g_{2} &=& \Gamma_8 g_1\Gamma_8.
\eea
In the following, without loss of generality, we work with the choice of parameters given by (\ref{two}).\par
The antisymmetry of $R_{IJ}$ for $I\neq J$ implies that there are at most $28$ generators defined by (\ref{R}).
One can nevertheless verify the existence, at the selected $c=\pm\frac{1}{12}$ values, of $7$ linear constraints satisfied by the $R_{IJ}$'s. This results in a total number of $21$ linearly independent generators. This is the number of $R$-symmetry generators entering the $F(4)$ superalgebra (whose $R$-symmetry is $so(7)$). \par
Within the $c=-\frac{1}{12}$ choice the seven constraints aree covariantly expressed as
\bea
2R_{i8} +C_{ijk}R_{jk}&=& 0.
\eea
As a consequence of these relations, the  $21$ linearly independent generators can be accommodated into the rank $2$ antisymmetric 
tensor $R_{ij}$. \par
We explicitly present for $c=-\frac{1}{12}$, in a covariant form, the non-vanishing \\ (anti)commutators involving  
the $R$-symmetry generators $R_{ij}$. We have
\bea\label{rq}
\relax [ R_{ij}, Q_k] &=& -\frac{i}{3}C_{ijk} Q_8+\frac{i}{3} C_{ijkl}Q_l+i\delta_{ik}Q_j -i\delta_{jk}Q_i,\nonumber\\
\relax[ R_{ij}, Q_8]&=& \frac{i}{3} C_{ijk}Q_k
\eea
and, similarly,
\bea\label{rqtilde}
\relax [ R_{ij}, {\widetilde Q}_k] &=& -\frac{i}{3}C_{ijk} {\widetilde Q}_8+\frac{i}{3} C_{ijkl}{\widetilde Q}_l+i\delta_{ik}{\widetilde Q}_j -i\delta_{jk}{\widetilde Q}_i,\nonumber\\
\relax[ R_{ij}, {\widetilde Q}_8]&=& \frac{i}{3} C_{ijk}{\widetilde Q}_k.
\eea
Furthermore, after lengthy but straightforward computations, the $R$-symmetry commutators can be covariantly expressed as
\bea\label{rr}
\relax[ R_{ij}, R_{kl}]&=& i(\delta_{ik}R_{jl}-\delta_{il}R_{jk}-\delta_{jk}R_{il}+\delta_{jl}R_{ik}) +\nonumber\\
&& -\frac{i}{6}(\delta_{ik}C_{jlmn}R_{mn}-\delta_{il}C_{jkmn}R_{mn}-\delta_{jk}C_{ilmn}R_{mn}+\delta_{jl}C_{ikmn}R_{mn})+\nonumber\\
&&\frac{i}{3}(C_{ijkm}R_{ml}-C_{ijlm}R_{mk}-C_{klim}R_{mj}+C_{kljm}R_{mi})+\frac{i}{3}(C_{ijm}C_{kln}R_{mn}).\nonumber\\&&
\eea
Despite the appearance in the right hand side of (\ref{rq},\ref{rqtilde},\ref{rr}) of the rank $3$ and rank $4$ totally antisymmetric octonionic structure constants,
the graded Jacobi identities are satisfied by construction. \par
The overall result here presented is a covariant realization, in terms of first-order and second-order matrix differential operators,
of the exceptional Lie superalgebra $F(4)$, whose even sector is given by $sl(2)\oplus so(7)$ and whose odd sector
contains $2\times8=16$ odd generators. The operators $H,D,K$ close the $sl(2)$ subalgebra, with the dilatation operator $D$ being the Cartan's element. The $so(7)$ subalgebra is realized by the $R_{ij}$ generators, while
the odd sector is given by the operators $Q_I$'s and ${\widetilde Q}_I$'s.

\section{The exceptional $F(4)$ deformed oscillator}

In Section {\bf 4} we proved that the scale-invariant  ${\cal N}=8$ supersymmetric Hamiltonians $H $ introduced in Section {\bf 3} become superconformal (with an $F(4)$ dynamical symmetry)  at the critical values $c=\pm\frac{1}{12}$. The resulting superconformal Hamiltonian, up to similarity transformations, is unique.\par
We introduce now the analogue of the De Alfaro-Fubini-Furlan construction \cite{dff}, presenting the (deformed) quantum oscillator possessing the exceptional Lie superalgebra $F(4)$ as its dynamical symmetry.  The deformed oscillator Hamiltonian is given by the linear combination $H+K$, where $K$ is introduced in (\ref{K}). 
$K$ adds the oscillator damping term to the Calogero potentials.\par
It is particularly rewarding to investigate the properties of the $H+K$ deformed oscillator Hamiltonian because it
possesses  a (degenerate) ground state with a discrete, bounded from below,  spectrum.\par
Both Hamiltonians, $H$ and $H+K$, can be compactly written as
\bea\label{scaham}
H_{\epsilon} &=&-\frac{1}{2}\partial_x^2{\mathbb I}_{16} +\frac{1}{2}\epsilon x^2{\mathbb I}_{16} +\frac{1}{x^2} V,\quad V=diag(v_1,v_2,\ldots, v_{16}),
\eea
for
\bea\label{criticalv}
&\quad  v_1=\frac{91}{72},\quad v_2= \ldots=v_8= -\frac{5}{72}, \quad v_9=\ldots =v_{16} = \frac{7}{72},&
\eea 
in such a way that $H$ is recovered for $\epsilon=0$ and the deformed oscillator Hamiltonian $H+K$ is recovered
for $\epsilon=1$.\par
Eight pairs of creation/annihilation operators ($I=,1,2,\ldots, 7,8)$ are introduced through the positions
\bea
a_I = Q_I -i{\widetilde Q}_I, && a_I^\dagger =Q_I+i{\widetilde Q}_I.
\eea
For later convenience we also introduce the symbol ${\overline I}=0,1,\ldots, 7$, identifying \\
$a_0:=a_8$, $a_0^\dagger :=a_8^\dagger$ and $\Gamma_0:=\Gamma_8$.\par
The Hamiltonian $H_{\epsilon=1}$ is recovered, for any ${\overline I}$, from the anticommutators
\bea
\frac{1}{2}\{a_{\overline I}, a_{\overline I}^\dagger\} &=& H_{\epsilon=1}.
\eea
The commutators give
\bea
\relax [a_{\overline I}, a_{\overline I}^\dagger] &=& {\mathbb I}_{16} +Y_{\overline I},
\eea
where $Y_{\overline I}=-[E_{\overline I},\Gamma_{\overline I}\Gamma_9]$ are constant and traceless diagonal matrices. We have
\bea
&Y_{\overline I} = diag(y_1,\ldots ,y_{16}),&\nonumber\\
&y_1 =-\frac{7}{3}, \quad y_2=\ldots=y_8=\frac{1}{3},\quad y_{9+{\overline J}}= -\frac{1}{3}+\frac{8}{3}\delta_{{\overline I}{\overline J}}.&
\eea
$Y_{\overline I}$ anticommutes with both $a_{\overline I}$ and $a_{\overline I}^\dagger$:
\bea
&\{a_{\overline I},Y_{\overline I}\}=\{a_{\overline I}^\dagger,Y_{\overline I}\}=0&.
\eea
Due to these properties, $a_{\overline I}$ ($a_{\overline I}^\dagger$) are annihilation (creation) operators such that
\bea
[H_{\epsilon=1}, a_{\overline I}] = -a_{\overline I}, &&[H_{\epsilon=1}, a_{\overline I}^\dagger] = a_{\overline I}^\dagger.
\eea
Each ${\overline I}$ defines $16$ lowest weight representations $|\lambda_k^{({\overline I})}>$ ($k=1,2,\ldots, 16$) 
introduced by the condition
\bea a_{\overline I}|\lambda_k^{({\overline I})}>&=&0.
\eea
Each $ |\lambda_k^{({\overline I})}>$ lowest weight vector is a $16$-dimensional vector whose only non-vanishing
component is in the $k$-th position. \par
We recall that a vector is bosonic (fermionic) if it is an eigenvector of the Fermion Parity Operator $\Gamma_9$ with eigenvalue
$+1$ ($-1$).
It is straightforward to show that the bosonic lowest weight vectors $ |\lambda_k^{({\overline I})}>$, which are obtained for $k=1,2,\ldots, 8$, do not depend on ${\overline I}$. They are annihilated by all eight operators $a_{\overline I}$. The lowest weight bosonic wave functions are proportional to
\bea\label{boslambda}
|\lambda_1^{({\overline I})}> &\propto& (x^{-7/6}e^{-\frac{1}{2}x^2},0,\ldots, 0)^T,\nonumber\\
|\lambda_j^{({\overline I})}> &\propto& (\delta_{jr}x^{1/6}e^{-\frac{1}{2}x^2})^T\quad (j=2,\ldots, 8; \quad r=1,\ldots, 16).
\eea
The picture is quite different for the fermionic lowest weight vectors $ |\lambda_k^{({\overline I})}>$ with $k=9,10,\ldots, 16$. None of them is annihilated by all eight operators $a_{\overline I}$.\par
For ${\overline I}=0,1,\ldots,7$, the fermionic lowest weight vector $|\lambda_{9+{\overline I}}^{({\overline I})}>$,  proportional to,
\bea\label{iifer}
&|\lambda_{9+{\overline I}}^{({\overline I})}> \propto (\delta_{{9+{\overline I},r}}x^{7/6}e^{-\frac{1}{2}x^2})^T
\eea
is only annihilated by the operator $a_{\overline I}$.\par
The fermionic lowest weight vectors $|\lambda_{9+{\overline I}}^{({\overline J})}>$, with ${\overline J}\neq {\overline I}$, proportional to
\bea\label{jneqi}
&|\lambda_{9+{\overline I}}^{({\overline J})}> \propto (\delta_{{9+{\overline I},r}}x^{-1/6}e^{-\frac{1}{2}x^2})^T,\quad ({\overline J}\neq {\overline I}),
\eea
are annihilated by the seven operators $a_{\overline J}$, ${\overline J}\neq {\overline I}$, while 
$a_I|\lambda_{9+{\overline I}}^{({\overline J})}>\neq 0$. It follows from (\ref{jneqi}) that 
$|\lambda_{9+{\overline I}}^{({{\overline J}_1})}>= |\lambda_{9+{\overline I}}^{({{\overline J}_2})}>$ for any pair ${\overline J}_1,{\overline J}_2\neq {\overline I}$.
\par
We have determined a total number of $24$ ($8$ bosonic and $16$ fermionic) different lowest weight vectors.\par
Their lowest weight energy $E_k^{({\overline I})}$ is computed from the relation
\bea 
E_k^{({\overline I})}&=& \frac{1}{2}<\lambda_k^{({\overline I})}|\{a_{\overline I}, a_{\overline I}^\dagger\} |\lambda_k^{({\overline I})}>=\frac{1}{2}<\lambda_k^{({\overline I})}|[a_{\overline I}, a_{\overline I}^\dagger] |\lambda_k^{({\overline I})}>= \frac{1}{2}+\frac{1}{2}<\lambda_k^{({\overline I})}|Y_{\overline I}|\lambda_k^{({\overline I})}>.\nonumber\\&&
\eea
We get
\bea\label{energy}
E^{({\overline I})}_1&=& -\frac{2}{3},\nonumber\\
E^{({\overline I})}_k&=& ~~\frac{2}{3}, \quad \quad\quad\quad \quad~ k=2,3,\ldots, 8,\nonumber\\
E^{({\overline I})}_{9+{\overline J}} &=& ~~\frac{1}{3} +\frac{4}{3}\delta_{{\overline J}{\overline I}}, \quad \quad
{\overline J}=0,1,\ldots, 7.
\eea
We will see in the following the implication of these results for the construction of the Hilbert space associated with the $F(4)$ deformed oscillator.\par
The eight creation (annihilation) operators $a_{\overline I}^\dagger$ ($a_{\overline I}$) are all unitarily equivalent. This is implied by the existence of seven unitary matrices $U_i$ ($U_iU_i^\dagger=U_iU_i^\dagger={\mathbb I}_{16}$)
satisfying
\bea\label{sevenu}
\relax &U_i\Gamma_iU_i^\dagger = \Gamma_8,\quad U_iE_iU_i^\dagger = E_8,\quad U_iY_iU_i^\dagger = Y_8,\quad [U_i,V]=0.&
\eea
In the above relations the repeated indices are not summed.\par
As a consequence of (\ref{sevenu}) we have, in particular,
\bea\label{unaadagger}
U_i a_i U_i^\dagger = a_0,&\quad &U_ia_i^\dagger U_i^\dagger = a_0^\dagger
\eea
for any $i=1,2,\ldots,7$.\par
An explicit expression for $U_1$ is given by
\bea
U_1 &=& E_{1,1}+E_{2,2}+E_{3,3}+E_{4,4}-E_{5,8}+E_{6,7}-E_{7,6}+E_{8,5}+\nonumber\\
&& E_{9,10}-E_{10,9}+E_{11,12}-E_{12,11}+E_{13,13}+E_{14,14}+E_{15,15}+E_{16,16},
\eea
where $E_{r,s}$ denotes the $16\times 16$ matrix with entry $1$ at the intersection of the $r$-th row with the $s$-th column and $0$ otherwise.\par
Similar expressions exist in the remaining cases ($i=2,\ldots, 7$). For simplicity they are not reported here.\par
As a corollary of the unitary relations (\ref{unaadagger}), the same set of lowest weight energy values
(one eigenvalue $\frac{2}{3}$, seven eigenvalues $-\frac{2}{3}$, seven eigenvalues $\frac{1}{3}$ and one eigenvalue
$\frac{5}{3}$) is encountered for each annihilation operator $a_{\overline I}$.
\section{Hilbert space and quasi-nonassociativity}
We determine the Hilbert space and the spectrum of the system with $F(4)$ as spectrum-generating superalgebra. \par
We point out that the bosonic lowest weight vector $|\lambda_1>\equiv|\lambda_1^{({\overline I})}>$ (the ${\overline I}$ dependence is dropped since the same lowest weight vector is shared by all ${\overline I}$'s)
of energy $-\frac{2}{3}$, see (\ref{energy}),  is not normalized as a square integrable function since its norm is negative. \par
Indeed the norm of the function $f(x) = x^{-\frac{7}{6}}e^{-\frac{1}{2}x^2}$ is computed as\par $(f,f)= \int_{-\infty}^{+\infty}|f(x)|^2dx =\int_{-\infty}^{0}|f(x)|^2dx+\int_{0}^{+\infty}|f(x)|^2dx$. \par
By setting $y=-x$ for $x<0$, we get $\int_{-\infty}^0|f(x)|^2dx= -\int_{\infty}^0|{(-y)}^{-\frac{7}{6}}e^{-\frac{1}{2}}y^2|^2dy = |(-1)^{-\frac{7}{6}}|^2\int_0^\infty y^{-\frac{7}{3}}e^{-y^2}dy$, so that
$(f,f) = (|(-1)^{-\frac{7}{6}}|^2+1)\int_0^\infty x^{-\frac{7}{3}}e^{-x^2}dx$. \par
After the $t=x^2$ change of variable we get
\bea
(f,f) &=& \frac{1}{2}(|(-1)^{-\frac{7}{6}}|^2+1)\int_0^{+\infty}t^{-\frac{5}{3}}e^{-t}dt = \frac{1}{2}(|(-1)^{-\frac{7}{6}}|^2+1)\Gamma(-\frac{2}{3}) <0.
\eea
The eight independent fermionic lowest weight vectors resulting from $ |\lambda_{9+{\overline I}}^{({\overline J})}>$, with ${\overline J}\neq {\overline I}$, have energy eigenvalue $+\frac{1}{3}=-\frac{2}{3}+1$. They correspond to the first excited states $a_{\overline I}^\dagger |\lambda_1>$. Contrary to
$|\lambda_1>$, their norm is positive.  Nevertheless, they have to be excluded from a Hilbert space since, by applying the annihilation operator $a_{\overline I}$, we obtain the state $|\lambda_1>$ with negative norm:
$a_{\overline I}a_{\overline I}^\dagger|\lambda_1> \propto|\lambda_1>$.\par
The lowest weight representation defined by $|\lambda_1>$ (which also includes the $ |\lambda_{9+{\overline I}}^{({\overline J})}>$, with ${\overline J}\neq {\overline I}$, vectors) does not define a Hilbert space because not all states are correctly normalized.\par
The Hilbert space induced by the $F(4)$ oscillator model is obtained by the direct sum of the remaining lowest weight representations.\par
It is convenient to rename as $b_i$ ($i=1,2,\ldots, 7$) the seven bosonic lowest weight vectors 
$\lambda_{i+1}\equiv \lambda_{i+1}^{\overline I}$ from (\ref{boslambda}) with positive energy eigenvalue $\frac{2}{3}$.  The eight fermionic lowest weight vectors $|\lambda_{9+{\overline I}}^{({\overline I})}>$ from (\ref{iifer}), with energy eigenvalue $\frac{5}{3}=\frac{2}{3}+1$, are the first excited states obtained from the 
$b_i$'s bosonic states. 
\par
The construction goes as follows.  We note at first that the creation operators $a_{\overline I}^\dagger$ satisfy the superalgebra
\bea\label{softsusy}
\{a_{\overline I}^\dagger, a_{\overline J}^\dagger \}&=& 2\delta_{{\overline I}{\overline J}} Z, \quad\nonumber\\
\relax [Z, a_{\overline I}^\dagger]&=&0,
\eea
which is formally equivalent to the ${\cal N}=8$ superalgebra (\ref{sqm}) of the Supersymmetric Quantum Mechanics. It should be noted however that the operator $Z$, defined by the first equation of ({\ref{softsusy}), is not the Hamiltonian, but a raising operator. The superalgebra (\ref{softsusy}) is an implementation of the concept of ``soft supersymmetry" discussed in \cite{cht}. \par
It is convenient to set $f_i\equiv |\lambda_{9+{i}}^{({i})}>$ and $f_0\equiv |\lambda_{9 }^{{(0)}}>$, so that we obtain
\bea
f_i&=& a_0^\dagger b_i,\nonumber\\
f_0&=&-a_1^\dagger b_1=-a_2^\dagger b_2=\ldots = -a_7^\dagger b_7
\eea
(for any $i$, $f_0=-a_i^\dagger b_i$).\par
The covariant relation
\bea
a_i^\dagger b_j &=& C_{ijk}f_k=C_{ijk}a_0^\dagger b_k
\eea 
is satisfied.\par
The vector space spanned by the direct sum of the seven lowest weight representations induced by the $b_i$'s vectors admits, by construction, normalized vectors. This Hilbert space corresponds to ${\cal L}^2({\mathbb R})^{16}$,
a $16$-ple of ${\cal L}^2({\mathbb R})$, the square-integrable functions on the real line. \par
Let us introduce the state $g_0$, given by
\bea
g_0 &:= a_0^\dagger f_0.
\eea
Its energy eigenvalue is $\frac{8}{3}$. We collectively denote as $w_r$, $r=1,2,\ldots, 16$
(by setting $w_i=b_i$, $w_8=f_0$, $w_{8+i}=f_i$, $w_{16}=g_0$), the $(7;8;1)$ states
$b_i; f_0,f_i;g_0$ of energy eigenvalues $(\frac{2}{3};\frac{5}{3};\frac{8}{3})$, respectively.
Let $E_r$ denotes the energy eigenvalue of $w_r$.
The Hilbert space of the theory is spanned by the states
\bea
&Z^nw_r, \quad r=1,\ldots ,16, \quad n\in {\mathbb N}_0,& 
\eea
where the operator $Z$ has been introduced in (\ref{softsusy}).\par 
The energy eigenvalue of $Z^nw_r$ is
\bea
E_{n,r} &=& 2n+E_r.
\eea
The energy spectrum of the theory is
\bea
&\frac{2}{3},\frac{5}{3},\frac{8}{3},\frac{11}{3},\ldots,&
\eea
Apart the ground state (which is $7$ times degenerate), each excited level is $8$ times degenerate:
\bea
&(7,8,8,8,\ldots )&
\eea
The degenerate states of each energy level are accommodated into a representation of the $so(7)$
$R$-symmetry subalgebra.\par
The operator $Z$ is introduced in (\ref{softsusy}) as square of the $a_{\overline I}^\dagger$ operators. 
For this reason the knowedge of 
the action of the $a_{\overline I}^\dagger$ operators on the $(7;8;1)$ states $w_r$'s is sufficient to reconstruct the whole semi-infinite tower of $(7,8,8,8,\ldots )$ states. The states $w_r$'s are accommodated
into the $(7;8;1)$ supermultiplet of the ${\cal N}=8$ worldline supersymmetry, see \cite{pato}. This supermultiplet (contrary to, e.g., the ${\cal N}=8$ supermultiplets $(k;8;8-k)$ with $k=2,3,4,5,6$)
preserves the octonionic covariance of the ${\cal N}=8$ worldline supersymmetry, see \cite{kuroto}.\par
The deformed (via Calogero potential terms) quantum oscillator induced by the exceptional superconformal algebra $F(4)$ is a unique system (up to similarity transformations) which only exists at a critical value of the parameters. It can be characterized by the property of ``quasi-nonassociativity".
The meaning is that the model is determined by the octonionic structure constants. Indeed, 
the structure constants of the $F(4)$ dynamical symmetry superalgebra are expressed in terms of the
rank $3$ and rank $4$ totally antisymmetric  octonionic structure constants, see formulas (\ref{rq},\ref{rqtilde},\ref{rr}). More than that, the ``strange" rational coupling constants of the $\frac{1}{x^2}$ Calogero potential term in the superconformal Hamiltonian (\ref{scaham}) are given by the diagonal matrix 
(\ref{criticalv}), which reads as
\bea
V=&diag(\frac{91}{72}, \frac{-5}{72},\frac{-5}{72},\frac{-5}{72},\frac{-5}{72},
\frac{-5}{72},\frac{-5}{72},\frac{-5}{72}, \frac{7}{72}, \frac{7}{72},\frac{7}{72},\frac{7}{72},\frac{7}{72},\frac{7}{72},\frac{7}{72},\frac{7}{72}).&
\eea
The above values are derived from  the octonionic structure constants $C_{ijk}$ through the formula
\bea\label{nice}
V
&=& \frac{1}{72}(\Gamma_8-\frac{1}{36}C_{ijk}\Gamma_i\Gamma_j\Gamma_k)C_{lmn}\Gamma_l\Gamma_m\Gamma_n.
\eea
The non-associativity of the octonions is encoded in the coupling constants of the $F(4)$ quantum oscillator. \par
It should be noted that the supertrace of $V$ is vanishing due to the relation
\bea
str V &=&\frac{1}{72}(91+7\times (-5) - 8\times 7)=0.
\eea 

\section{Conclusions}

The Hilbert space of the one-dimensional $F(4)$ deformed quantum oscillator is given by a $16$-ple of square-integrable functions. The energy levels are quantized to be $\frac{2}{3} +{\mathbb N}_0$. The ground state of energy $\frac{2}{3}$ is $7$ times degenerate, while the excited states are $8$ times degenerate. The eight creation
operators $a_{\overline I}^\dagger$ close the ${\cal N}=8$ worldline superalgebra (\ref{softsusy}).\par
Therefore, the states of the seminfinite tower $(7,8,8,8,\ldots )$ are interconnected by the worldline supersymmetry induced by the irreducible, see \cite{pato}, $(7;8;1)$ supermultiplet.\par
The model is uniquely determined (up to similarity transformations). The associated superconformal quantum mechanics is obtained from the most general octonionic-covariant and scale-invariant ${\cal N}=8$ supersymmetric quantum mechanics at a critical value of enhanced $F(4)$ dynamical symmetry.\par
The non-trivial Calogero's coupling constants are expressed in terms of the octonionic structure constants.\par
It is worth mentioning that, besides the construction of the $F(4)$ model, other results have been obtained by
our investigation using the octonionic-covariant approach. We excluded, at ${\cal N}$=7, the existence of an octonionic covariant superconformal quantum mechanics realized on $16$ fields and  with $G(3)$ as spectrum-generating superalgebra. Even if a priori possible, not such a model exists for any combination of the parameters allowed by octonionic-covariance.\par 
The superalgebra $D(4,1)\approx osp(8|2)$ is the spectrum-generating superalgebra of the undeformed system (all Calogero's coupling constants are set to zero) given by the direct sum of $16$ ordinary one-dimensional oscillators. The details of the construction will be reported in a larger forthcoming paper devoted to a general derivation of spectrum-generating superalgebras of (un)deformed quantum oscillators.\par
\par
Introducing nonassociativity in physics and quantum mechanics is notoriously a tricky business. So far, the unique truly nonassociative quantum mechanical system we are aware of, based on the Jordan's formulation of quantum mechanics, was derived in 1934 by Jordan, von Neumann and Wigner \cite{jnw} and further analyzed in \cite{gpr}. The observables are accommodated in the $3\times 3$ Hermitian octonionic matrices (the exceptional Albert algebra). \par
There are several examples where nonassociative structures are relegated into the non-observable sector of
a quantum theory.  For instance, the associativity deficit of the twisted cocycle condition of quasi-Hopf algebra, defined in \cite{maj} and discussed in \cite{mss}, is often realized in canonical quantization, see \cite{bl}, as a phase when projected onto the Hilbert space of the theory. The phase is a unitary transformation which can be reabsorbed in the normalized ray vector.\par
A different type of nonassociativity is encountered in the $F(4)$ deformed quantum oscillator model here presented. The nonassociativity of the octonions is encoded, as ``quasi-nonassociativity", in the Calogero coupling constants of the theory which are given, see formula (\ref{nice}), in terms of the octonionic structure
constants. The existence of such a relation is a consequence of the model admitting an octonionic-covariant formulation.
\\ {~}~
\par {\Large{\bf Acknowledgments}}
{}~\par{}~\par
Z.K. and F. T. are grateful to the Osaka Prefecture University for hospitality. 
This research was supported by CNPq under PQ Grant 306333/2013-9.
N. A. is supported by the  grants-in-aid from JSPS (Contract No. 26400209).

\end{document}